\documentclass[12pt,preprint]{aastex}
\bibpunct{(}{)}{;}{a}{}{,} 

\shortauthors{Li et al.} \shorttitle{Resolving the nature of the
Rosette HH2 jet}

\begin{document}

\title{The dissolving Rosette HH2 jet bathed in harsh UV radiation of the Rosette Nebula}
\author{J. Z. Li$^{1}$, Y. -H. Chu$^{2}$, R. A. Gruendl$^{2}$}
\affil{$^{1}$National Astronomical Observatories, Chinese Academy
of Sciences, 20A Datun Road, Chaoyang District, Beijing 100012, China; ljz@bao.ac.cn \\
       $^{2}$Department of Astronomy, University of Illinois, Urbana-Champaign, IL, 61801, USA\\
}

\begin{abstract}

Herbig-Haro flows discovered in photoionized medium forms a
separate category and detailed studies of which become one of the
key issues to our understanding of jet production and evolution.
The Rosette HH2 jet is the second of such flows that immersed in
the spectacular HII region of the Rosette Nebula. However, its
disconnected jet components are detached from the proposed energy
source, have additional unusual properties and thus a disputable
nature. In this paper, we investigate through high-quality echelle
spectrographs the physical nature of the jet system. The jet shows
distinctly different velocity components. It is believed to be
composed of a fast neutral jet with an approaching velocity of
-39.5 km $^{-1}$ as respect to the systemic rest frame, and likely
an extensive, photoevaporated envelope dissolving at roughly the
sound speed. This led us to infer a fast dissipating nature of the
jet system being bathed in the fully photoionized medium of
Rosette.

In addition, time series photometric observations provide evidence
that the energy source is highly variable, with amplitudes of up
to ${>}$ 1~mag in R \& I. This is consistent well with an early
evolutionary status of the jet driving star with a red, late type
spectrum in the optical.

\end{abstract}


\section{Introduction}

The recent discovery of an increasing number of collimated
Herbig-Haro (HH) jets ejected by optically visible, young low-mass
stars that immersed in HII regions makes them a separate category
of HH flows (Reipurth et al. 1998; Bally et al. 2000; Bally \&
Reipurth 2001; Li 2003; Li \& Rector 2004). Detailed
investigations of these so called photoionized jets are of
particular interest as they are in association with visible young
stellar objects at their early stages of evolution. These jet
driving sources would otherwise have been shielded by optically
opaque envelopes and/or natal molecular clouds if not made visible
by photoionization of the HII regions in which they reside. On the
other hand, the lifetime of such photoionizing jets can be short,
this makes them hard to identify and even more crucial in studies
of jet production and maintenance before the beams are lost in the
glare of the ionized nebulae.

Properties of the externally ionized jets are found to heavily
rely on local conditions of the photoionized medium i.e. the
intensity of the UV radiation field. Some jet systems such as the
Rosette jets (Li 2003; Li \& Rector, 2004) even contradict in
properties with conventional impressions of a jet. The Rosette HH2
jet, for example, is proposed by Li (2003) as a high-excitation
jet from a late type star around the central cavity of  Rosette
that shows an unusual appearance. It is composed of mainly two
discrete knots or jet components probably in the process of
photo-dissipation. These components resemble more nebulous
entities, the proposed jet therefore shows an abnormally large
width to length ratio which is in remarkable discrepancy from
conventional impressions of a jet. Furthermore, the seemingly
collimated part of the jet shows a detached appearance from both
the suggested energy source and the extensive portion of the jet.
This jet system displaying various anomalous features, if finally
convinced, will be the only other case of a high-excitation jet
that survived the harsh UV ionization of the Rosette Nebula.
Numerous young stellar objects originated from the same episode of
star formation in this region could have already shed their
envelopes, ceased their mass ejection and the signatures of the
existence of any flows were also dissipated by the external
ionization. These two jet systems may well be the last two still
identifiable and are also in the process of their fast
photo-dissipation. Detailed studies of such jets will definitely
contribute to a better understanding of the interactions between
fierce external UV ionization and the jet systems, and to the
final theoretical solution to jet formation and evolution and
their dependence on environmental conditions. There are, however,
still possibilities of solely spatial coincidence of
high-excitation gas entities in Rosette along the line of sight,
rather than being manifestations of discrete ejecta from the
proposed energy source as a result of episodic or unsteady mass
outflow? This causes confusion since its discovery and definitely
awaits for further clarification by high quality observational
studies. This study presents primarily the kinematics of the
proposed jet system, which definitely consolidated its physical
origin as a jet.





\section{Observations and data reduction}

\subsection{Echelle Spectroscopy}

Single-order echelle spectra, covering both H$\alpha$ and [NII],
of the jet system were obtained with the CTIO Blanco 4m telescope
and its echelle spectrograph at the Ritchey-Critchien focus on
January 12, 2005. A resolving power of $\sim$ 40000 was achieved
around H$\alpha$. The spectral data reduction were performed
following standard procedures in IRAF. The wavelength calibration
of the data has been improved by verifying the night sky lines,
which is expected to be accurate to $\pm$ 1 kms$^{-1}$. The final
echelle spectrogragh are corrected to the heliocentric rest frame.

\subsection{Photometric and spectroscopic monitoring}

We have initiated a simultaneous photometric and spectroscopic
monitoring campaign of the jet driving sources in Rosette between
Dec. 31, 2004 and Jan. 7, 2005. For a detailed description of the
monitoring campaign, please refer to Section 2.3 of Li et al.
(2006). Due to the faintness of the Rosette HH2 source, only two
consecutive spectra were achieved by the 2.16 m telescope of the
National Astronomical Observatory (NAOC) during this run of
observations. The Beijing Faint Object Spectrograph and Camera
(BFOSC) and a thinned back-illuminated Orbit 2k$\times$2k CCD were
used. The G4 grating of BFOSC resulted in a two-pixel spectral
resolution of 8.3~\AA. The differential photometric observations
based on the Hsing-Hua 80 cm telescope, which is located at the
Xing-Long station of NAOC, were later extended to Jan. 12, 2005.
Differential light variations of the jet source during this period
were obtained through both the R \& I band filters, resulting in
differential magnitudes accurate to within 0.04 mag for both
filters.

\section{Kinematics of the jet}

As mentioned in the introduction section, the proposed jet shows
many unique features and its physical nature has to be clarified.
The high-quality echelle spectrograph along the jet is shown by
Fig. 1a. It is clear that H$\alpha$ emission from the jet is
strong but has a broad appearance, which is severely blended with
the background nebular emission. Fortunately, the [NII] emission
lines have comparatively low dispersion in velocity and are
ideally resolved from at least those from the receding shell of
the HII region. Based on the well-resolved background emission in
the counterjet direction, the collisionally excited [NII] emission
lines from both the approaching and the receding shells of the
Nebula, indicate distinct heliocentric radial velocities
(V$_{hel}$) that centered at 0.3 \& 29~km~s$^{-1}$, respectively.
This gives a systematic expansion of 14.3 $\pm$ 1~km~s$^{-1}$ of
the ionized gas and a systemic V$_{hel}$ of 14.6 $\pm$
1~km~s$^{-1}$ at this part of the Rosette Nebula. It can be
noteworthy that the forbidden [NII] emission from the receding
side of the nebula is stronger, which hints for a larger emission
measure and indicates the approaching shell is shallower than the
receding part in the line of sight of the jet system.

Both the H$\alpha$ and [NII] emission from the jet, however,
indicate complex structures in the velocity field. To make things
clear, we present in Fig. 1b the echelle spectrograph after
careful background subtraction. In general, the blue wing of the
radial velocity (RV) distribution of the jet materials brakes
slowly along the jet direction and finally
merges with the red wing, indicating a constant RV, at the
distance of the extensive part of the jet with a diffuse
appearance.
The is distinctively illustrated by the Position-Velocity plots
shown by Figs. 2a \& 2b. The RV distribution of the jet materials
alone verifies that the discrete components of the jet proposed by
Li (2003) are indeed kinetically and physically related.



The [NII] emission lines at $\lambda\lambda$6548 \& 6583, however,
are further resolved into a high-velocity component (HVC) with a
heliocentric RV centered at -25~km~s$^{-1}$
(V$_{sys}$=-39.5~km~s$^{-1}$) and a low-velocity component (LVC)
with a heliocentric RV of 5~km~s$^{-1}$
(V$_{sys}$=-9.5~km~s$^{-1}$) in a distinct manner. This is
important as the existence of the HVC makes solid the jet nature
of the system, excluding possibilities of solely evaporated clumps
of gas happen to be projected in the close vicinity of the energy
source, which otherwise can not reconcile with the HVC and the two
fold velocity structures detected. Furthermore, the well-resolved
data permit us to distinguish between the physically related
outflowing components with distinct RV, which points to a clear
physical picture of the jet system. It is composed of (1) a HVC
that attributed to a fast neutral jet that remains propagating at
the original velocity into the photoionized medium. This HVC,
however, later begins to decelerate as the distance from the jet
source increases due to probably external UV destruction, and (2)
a LVC in association with gas entrained by the jet but keeps being
photoevaporated by the strong UV field of Rosette. It shows a RV
in good agreement with a fast dissipating envelope moving at the
sound speed of around 10 km s$^{-1}$ (Johnstone et al. 1998).
This, on the other hand, naturally explains why the jet has a
large width to length ratio or why the jet components resulted
from discrete mass ejection have in common an extended or even a
nebular appearance, which is particular to this jet in most likely
a dissolving phase. We would therefore feel safe to declare that
as evidenced by both the appearance and the kinematics of the
Rosette HH2 jet, we are spotting the latest stages of jet
evolution or dilution in a photoionized medium.







\section{Spectral properties of the jet system}




The extracted single-order echelle spectrum of the energy source
is presented by Fig. 3a, where H$\alpha$ is well resolved from the
ambient [NII] emission lines. The H$\alpha$ in emission with a
detected equivalent width of 6.9~\AA~ is broad as mentioned in the
above section and shows a complex profile. It has signatures of
absorption in both the blue and the red wings, indicating most
likely the co-existence of both mass outflow and inflow. This is
in agreement with the jet driving nature of the young stellar
object. The [NII] emission lines $\lambda\lambda$6548 \& 6583
indicate an equivalent width of 2.1~\AA~and 0.4~\AA, respectively.
This yields a line ratio of $\sim$ 5. The extracted spectrum of
the collimated part of the jet is given in Fig. 3b. Note about the
broad, multi-component emission associated with its H$\alpha$
emission and the double peak profile of the forbidden [NII]
emission lines, indicating the distinct existence of different
velocity components in the ejected materials.

Low-resolution spectrum of the Rosette HH2 source is presented in
Fig. 4, which illustrates primarily moderate H$\alpha$, prominent
[OIII] emission and shows a late spectral type with a very red
continuum in the optical. Low-resolution spectroscopy was employed
due to the faintness of the energy source in the optical at the
large distance of Rosette of 1.39 kpc (Hensberge et al. 2000). As
a result, the H$\alpha$ emission is hardly resolved from the
nearby [NII] emission lines. This already benefits from the
comparatively low thermal width of its H$\alpha$ and nearby
emission lines as compared to other active young stellar objects.
The equivalent width of H$\alpha$ varies from 12.2 to 20.5~\AA~ in
the two consecutive exposures. The stronger emission from
H$\alpha$ detected by the low-resolution spectroscopy is
attributed to more likely the highly variable nature of the jet
source rather than a significant contribution from the blended
[NII] emission lines. However, its high state of excitation as
revealed by the prominent [OIII] emission lines at
$\lambda\lambda$4959 \& 5007 strongly suggests a fully ionized
origin of at least the outer layers of the relic disk in
association. This will definitely result in a rapid
photodissipation of the system.


%

\section{Variability of the jet-driving source}

Results from the photometric monitoring in R and I were presented
by Figs. 5a \& 5b, respectively. It is clear that both lightcurves
show anomalously strong photometric variations in a largely
consistent way in both bands. This convinces us the large
amplitude variations are true in light of the small photometric
uncertainty in each band (please refer to section 2.2).
Furthermore, possibly two consecutive eruptive events with similar
amplitudes of $\sim$1.4 mag in R and 1.25 mag in I, respectively,
were detected by the 13 days monitoring campaign. Its large
amplitude of variation is commensurate with and set solid its
young status evolution of the energy source. The erratic light
variations are most likely attributed to prominent chromospheric
activity or rather unsteady mass accretion from at most a relic
disk (Li 2005).

\section{Summary}

Based on our high-spatial and high-spectral resolution
spectroscopy by the Blanco 4m telescope of CTIO, we presented for
the first time the kinematics of the Rosette HH2 jet, which
clarified also its physical nature of the jet system. The
collimated part of the jet shows clearly two distinct velocity
components i.e. a HVC with a RV of -39.5~km~s$^{-1}$ and a LVC
with a RV of -9.5~km~s$^{-1}$ as respect to the rest frame of
Rosette. We thus conclude that the jet is composed of a neutral
core that remains propagating at its original velocity into the
photoionized medium, and a fast dissipating envelope dissolving at
roughly the sound speed. This jet system, along with the Rosette
HH1 jet, is believed to be in a process of fast dissipation due to
the harsh UV radiation field in which it resides. The time series
photometric observations signifies a highly variable nature of the
jet driving source, in nice agreement with its young stage of
evolution and possibly an association with a relic disk also in
the fate of photoevaporation and dissipation.

{\flushleft \bf Acknowledgments~}

Thanks to Y. L. Qiu and J. Zh. Li for coordinating the photometric
observations with the Hsing-Hua 80 cm telescope. The author is
grateful to W. Su for her help on the data reduction. This project
is supported by the National Natural Science Foundation of China
through grant O611081001.


\clearpage

\figcaption[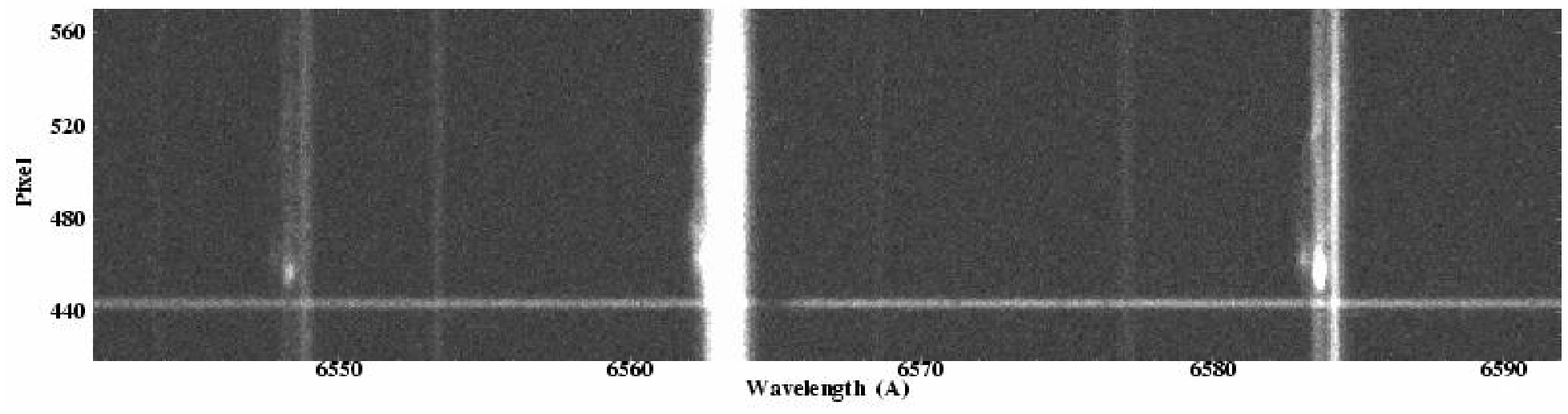,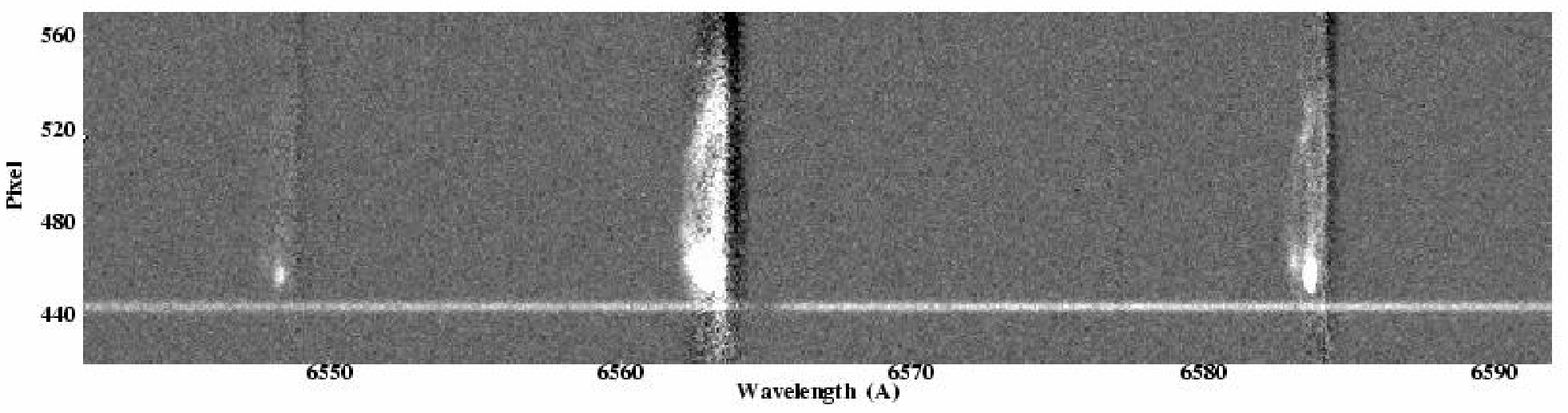]{Single-order echelle spectrograph
of the jet system covering both H$\alpha$ and [NII]. The slit is
oriented along the jet direction with a position angle of
315\arcdeg. Upper panel: Wavelength calibrated single-order
echelle spectrograph of the jet. Note that the jet is clearly
resolved into two distinct velocity components as disclosed by
especially the [NII] emission lines, which have a less thermal
dispersion than H$\alpha$. Line emission from both the receding
and the approaching shells of the HII region are well resolved and
presented. Lower panel: Net emission from the jet system after
careful background subtraction.}

\figcaption[fig2a.ps,fig2b.ps]{Position-Velocity diagrams of the
H$\alpha$ (left panel) and [NII] $\lambda\lambda$6583 emission
(right panel) from the jet system. The radial velocity in the
abscissa is calibrated to the heliocentric rest frame. The
ordinate indicates projected distance from the jet source, the
position of which is marked as zero in the plot.}

\figcaption[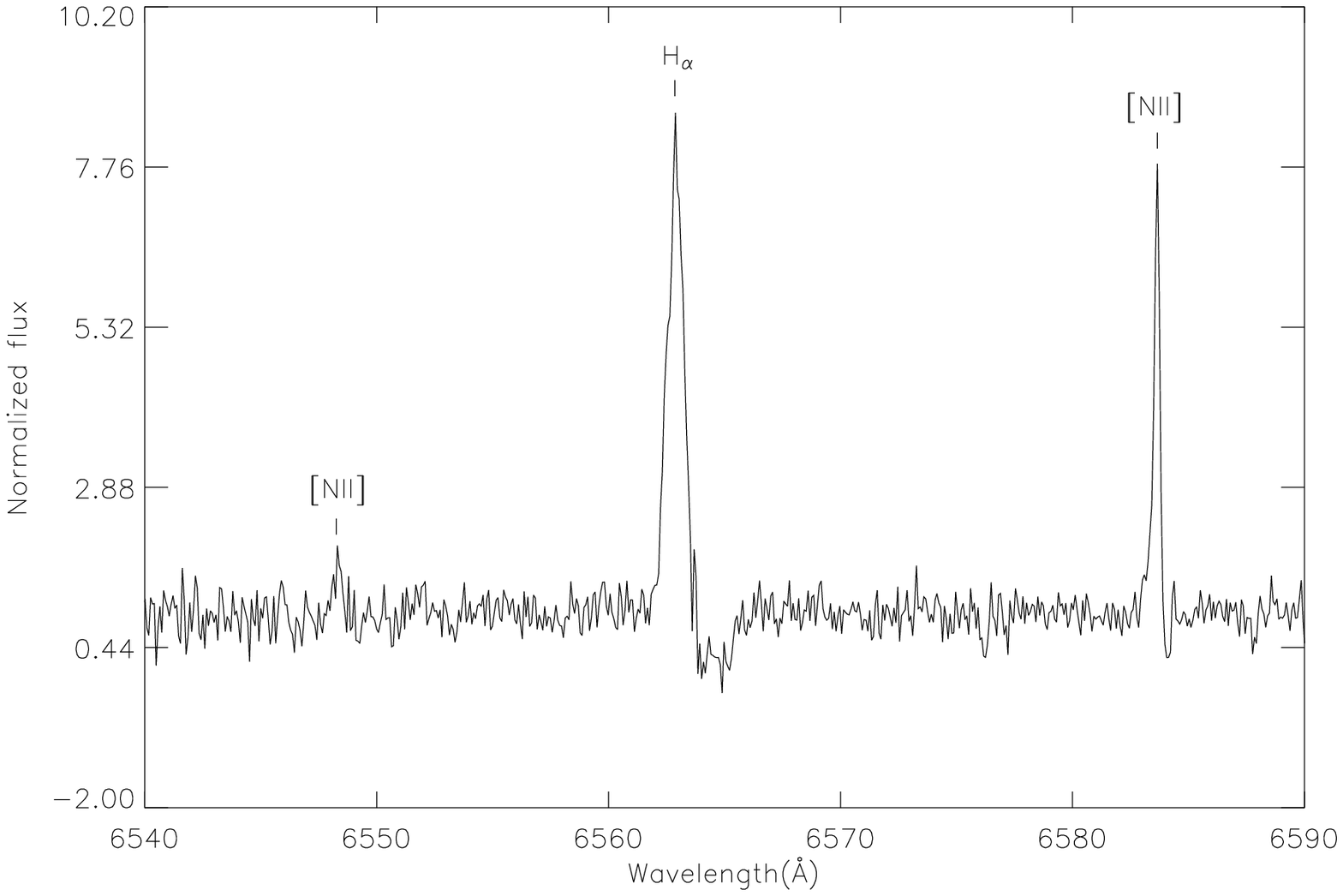,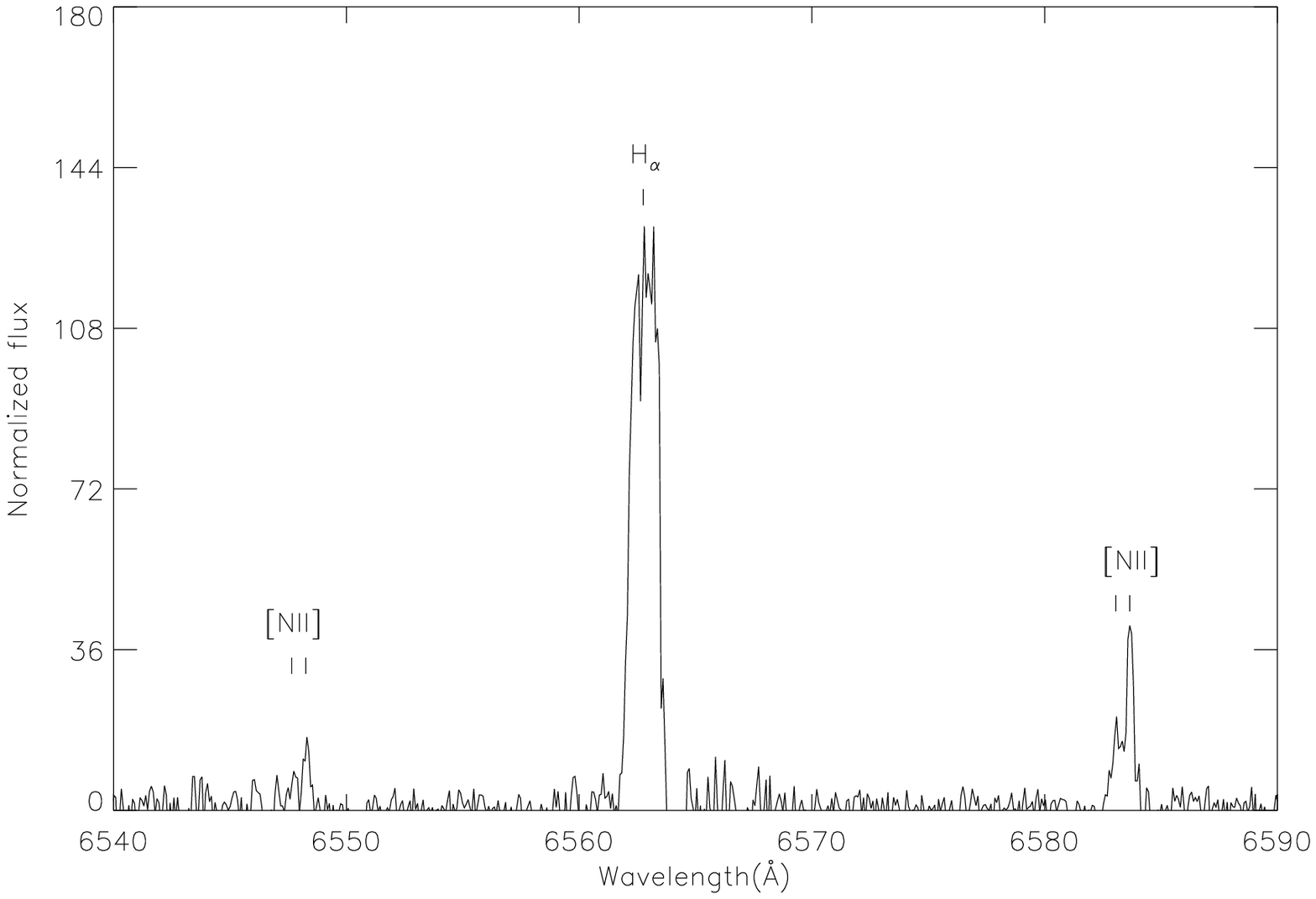]{Normalized echelle spectrum of the
jet-driving source (upper panel) and that of the jet (lower
panel). Note about the double peak profiles associated with both
of the [NII] emission lines.  The intensity ratio of the two [NII]
emission lines I($\lambda$6583)/I($\lambda$6548) in the source
spectrum equals 3.6 and that in the jet 4.5. The equivalent width
of H$\alpha$ emission of the jet at 10\% intensity amounts to
163~\AA, whilst that of the source spectrum is 5.8~\AA.}

\figcaption[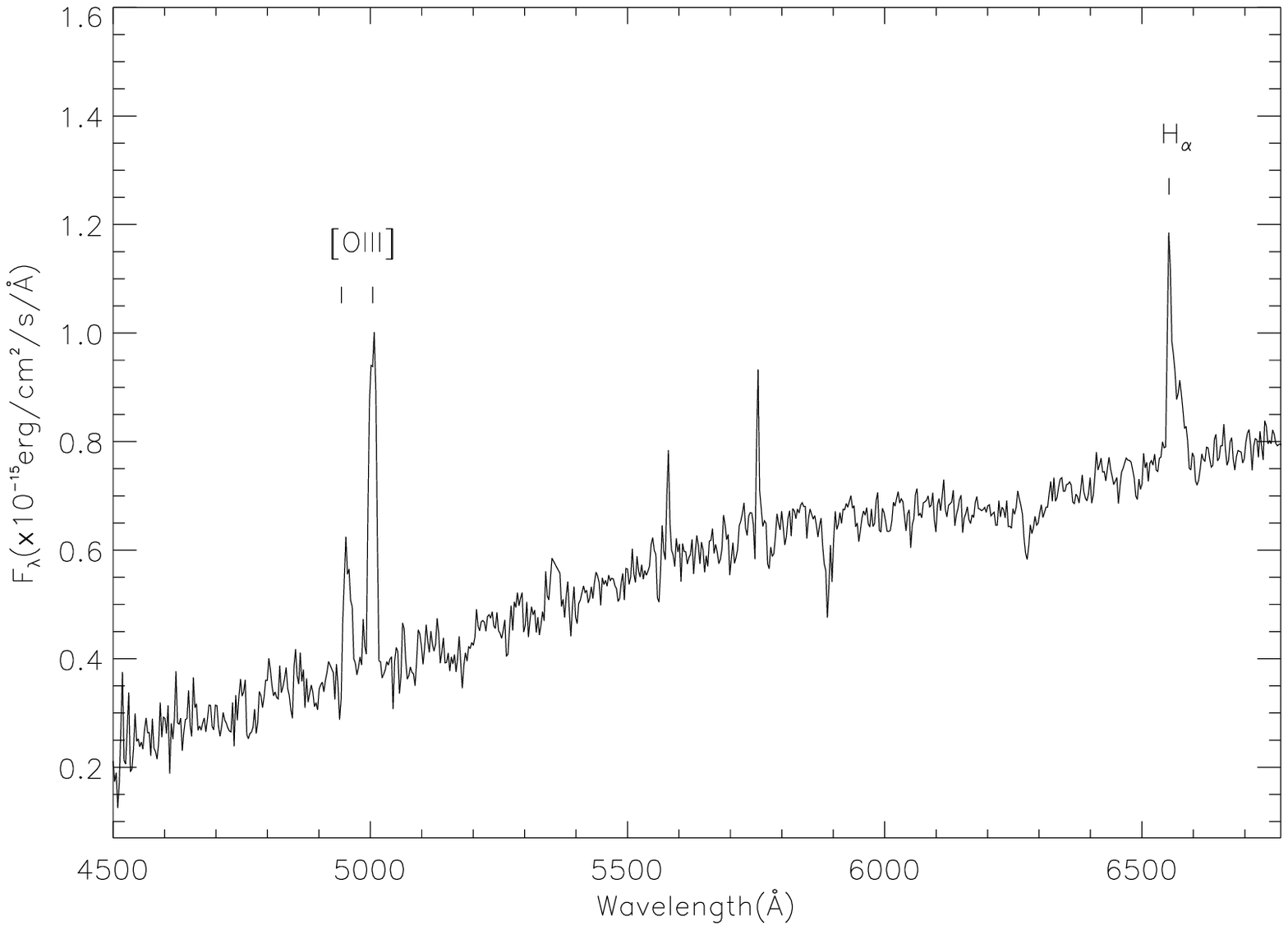]{Low-resolution spectrum of the jet source
covering both H$\alpha$ and the [OIII] emission lines. Note about
the red continuum of the energy source with a late spectral type.}

\figcaption[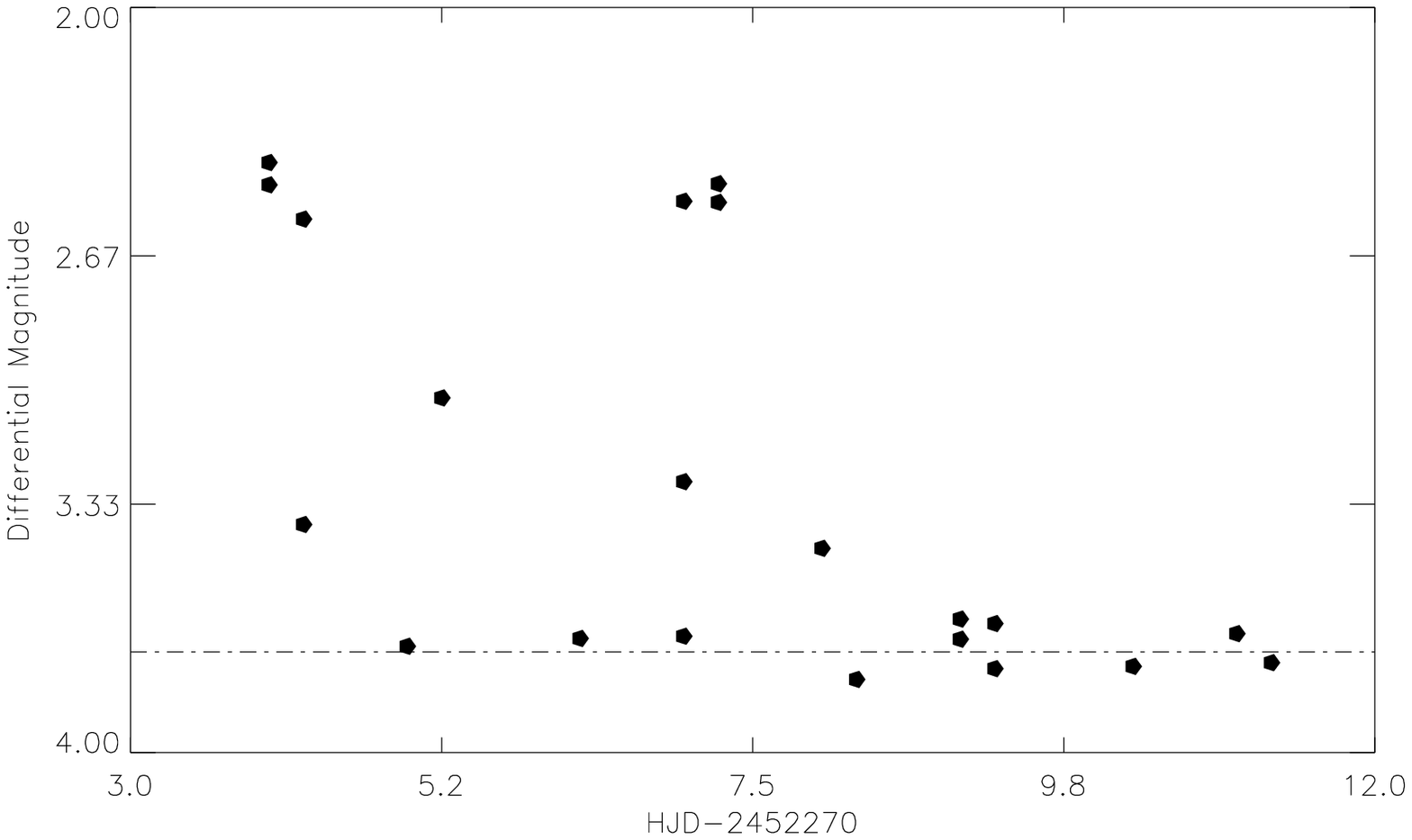,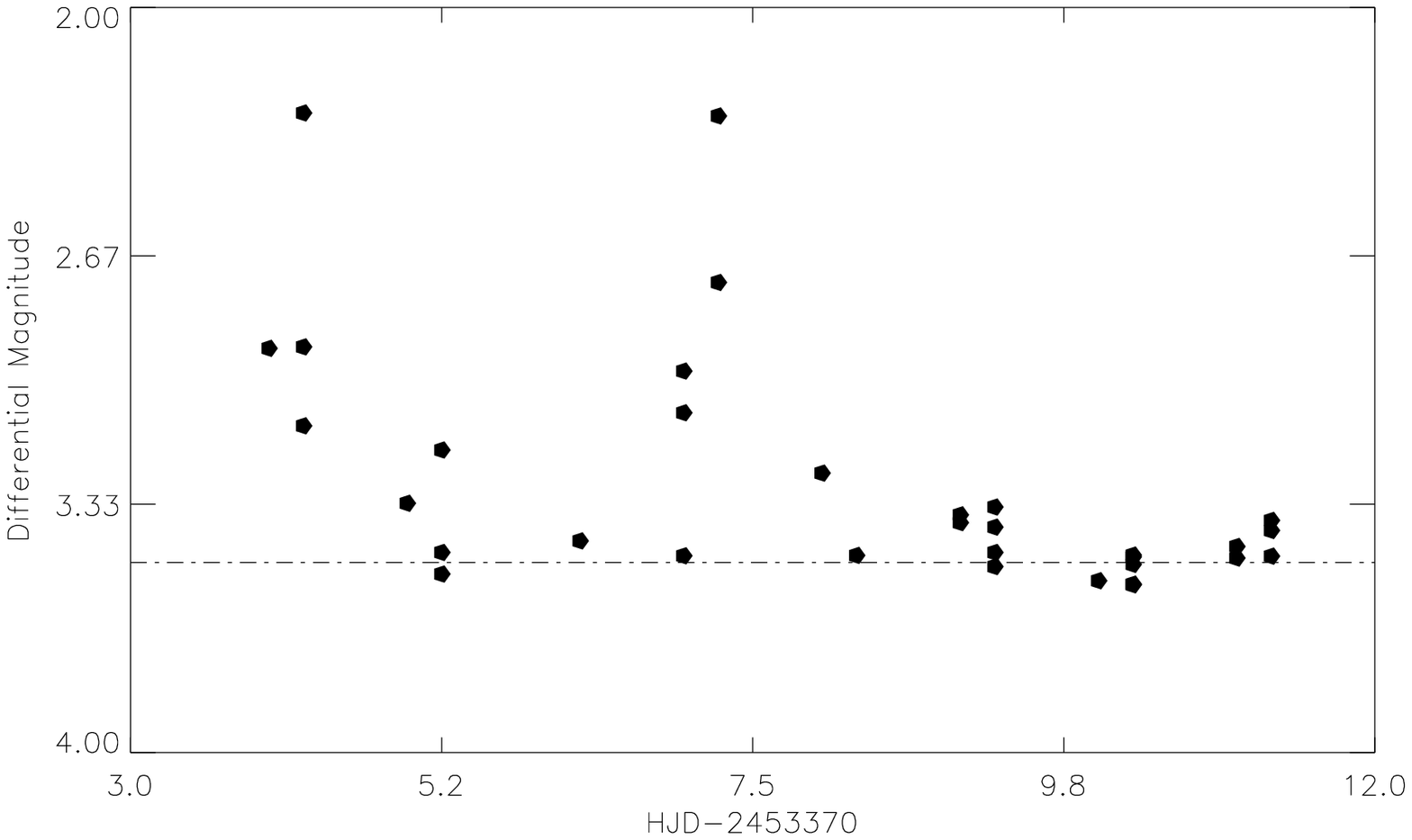]{Light variations of the jet source
in both R (upper panel) and I (lower panel). Note about the
detection of two flare-like events with a similar amplitude in
both filters. The dot-dashed line in each panel indicates a fitted
normal level of the brightness of the jet source with the large
amplitude variations cancelled.}

\clearpage
\plotone{fig1a.ps}
\plotone{fig1b.ps}
Fig.1
\clearpage
\epsscale{0.55}
\plotone{fig2a.ps}
\plotone{fig2b.ps}
Fig.2
\clearpage
\epsscale{0.8}
\plotone{fig3a.ps}
\plotone{fig3b.ps}
Fig.3
\clearpage
\epsscale{1.0}
\plotone{fig4.ps}
Fig.4
\clearpage
\epsscale{0.8}
\plotone{fig5a.ps}
\plotone{fig5b.ps}
Fig.5

\end{document}